# Process-oriented Iterative Multiple Alignment for Medical Process Mining

Shuhong Chen[1], Sen Yang[1], Moliang Zhou[1], Randall S. Burd[2], Ivan Marsic[1]
[1] Rutgers University, NJ, USA; [2] Children's National Medical Center, Washington, D.C., USA
[1] {sc1624, sy358, mz330, marsic}@rutgers.edu; [2] rburd@childrensnational.org

*Abstract*—Adapted from biological sequence alignment, trace alignment is a process mining technique used to visualize and analyze workflow data. Any analysis done with this method, however, is affected by the alignment quality. The best existing trace alignment techniques use progressive guide-trees to heuristically approximate the optimal alignment in $O(N^2L^2)$ time. These algorithms are heavily dependent on the selected guide-tree metric, often return sum-of-pairs-score-reducing errors that interfere with interpretation, and are computationally intensive for large datasets. To alleviate these issues, we propose process-oriented iterative multiple alignment (PIMA), which contains specialized optimizations to better handle workflow data. We demonstrate that PIMA is a flexible framework capable of achieving better sum-of-pairs score than existing trace alignment algorithms in only $O(NL^2)$ time. We applied PIMA to analyzing medical workflow data, showing how iterative alignment can better represent the data and facilitate the extraction of insights from data visualization.

## I. INTRODUCTION

Process mining can be used to extract insights from workflow data, and has been used in various fields like business and healthcare to analyze and improve different processes [4][6][14]. One process mining technique is trace alignment [4]. Alignment is useful in process mining because it allows intuitive comparisons between the activity sequences of a process's observed traces. This property makes alignment applicable to workflow visualization, pattern discovery, and deviation detection [4][6][10]. For example, trace alignment was used to analyze the diagnostics procedures of patients [2]. Another study derived insights from a consensus sequence alignment of the laparoscopic cholecystectomy workflow [5]. Alignment information has also proven useful in guiding workflow discovery [15].

A process is a set of activities that occur in a specific temporal order based on a set of rules. Each realization of a process, called a trace, may be observed and recorded as a sequence of activities. A collection of traces becomes a workflow log. Trace alignment, adapted from biological sequence alignment [4][9], takes a log and returns a two-dimensional alignment matrix. The matrix contains a row for each trace. In this matrix, the traces' activity sequences are preserved, but gaps are inserted between activities so that common activities are aligned across the columns. The resultant alignment matrix is said to have dimensions N by L, where N is the number of traces (in rows) and L is the alignment length (columns).

Errors in an alignment include having common activities spread across more columns than necessary, failing to align activities to columns of higher frequency, and incorrectly prioritizing alignment of infrequent activities [4]. For a given alignment, the magnitude of these errors is most commonly measured by sum-of-pairs score. Sum-of-pairs score is the sum of the Hamming distances between every pair of rows in an alignment matrix (Eq. 1).

$$SPS(M) = \sum_{i=1}^{N-1} \sum_{j=i+1}^{N} \sum_{k=1}^{L} I(M[i,k] = M[j,k]) \quad (1)$$

where **M** is the alignment matrix, with one trace per row, each trace aligned across the column. **N** is the number of traces and **L** is the longest trace length. **I** is the indicator that the activities at **M**[i,k] and **M**[j,k] match (1 if both gaps or both activities, 0 otherwise).

Sum-of-pairs score thus measures how much variation is present between the aligned traces: as more activities are spread out across different columns, higher Hamming distances will add up. Minimizing this sum-of-pairs score is the objective goal of alignment, and is the standard referred to in terms of performance. Finding the global sum-of-pairs minimum between more than two traces is an NP-hard problem, and takes $O(2^{NL})$ time. Algorithmic work tries to heuristically find alignment approximations in less time.

### A. Related Work

Multiple sequence alignment (MSA) was first introduced in bioinformatics, where it was used to find commonalities between biological sequences such as DNA and proteins [4][9][11]. The Needleman-Wunsch algorithm [9] is a basic approach for MSA which constructs a dynamic programming table using an objective function to compute a pair-wise alignment (step 6 in Figure 1). Over the last several decades, extensive work on optimizing MSA for use in the life sciences has been performed [12]. Biological sequence data, however, is different from workflow data in several ways: 1. process traces are more variable in length; 2. bio-sequences have known scores and penalties for substitutions and indels; and 3. DNA or proteins involve at most twenty or so unit types, while processes may have dozens of activity types to align [4].

These differences mean that many of the improvements to bioinformatics MSA do not apply to process mining. The current trace alignment algorithms [3][4][13] do not properly apply MSA to workflow analysis. Some work has been done to tailor MSA to process mining, such as the inclusion of dynamic time warping that incorporates activity duration information into alignment [13]. These previous approaches have not addressed the critical issue of inappropriate progressive guide-tree usage.

The progressive guide-tree approach is derived directly from biological MSA [12]. Because finding the optimal alignment between sequences takes non-polynomial time with respect to the number of sequences, only pair-wise alignment is feasible. In light of this limitation, the progressive guide-tree methods follow the underlying assumption that sequences and traces more similar to each other should be aligned first. All edit distances between pairs are used to hierarchically cluster the sequences. The resulting dendrogram is a guide-tree that dictates the order of pair merging [4].

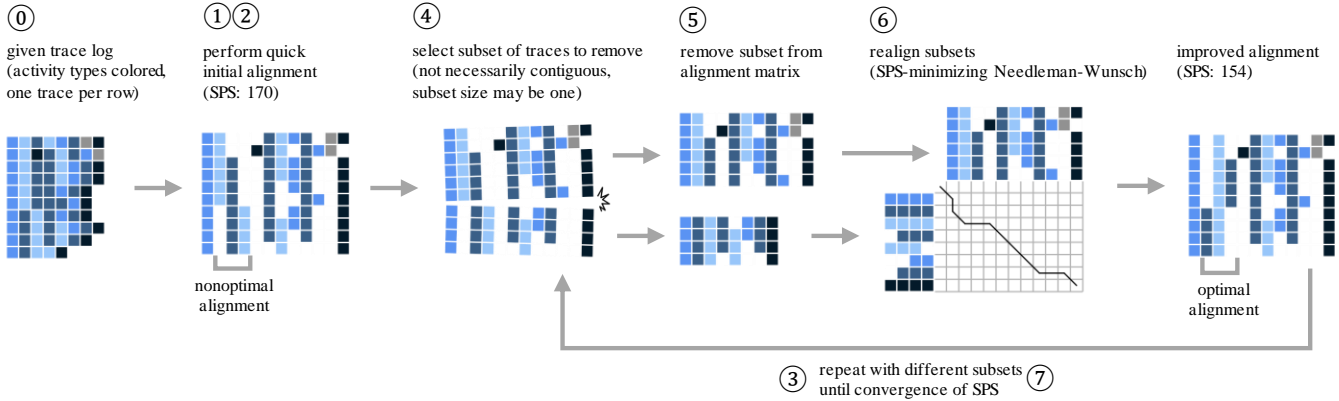

Figure 1. Framework of our Process-oriented Iterative Multiple Alignment (PIMA) algorithm. (Best viewed in color.)

*B. Problem Statement*

The foremost problem with progressive guide-tree alignment is the time complexity. Guide-tree construction requires $O(N^2L^2)$ time to calculate, as $O(L^2)$ must be spent to calculate the distance for each of the $O(N^2)$ trace pairs [3]. The quality of the final alignment is heavily dependent on the quality of the tree, so metrics less informative than $O(L^2)$ edit distance will negatively impact the results [4]. The $O(N^2)$ term is particularly disadvantageous for big-data process mining applications, where logs have many traces. Large values of N may cause very long running times (on the order of weeks). Reducing $O(N^2)$ to $O(N)$ significantly improves efficiency.

Another issue with the progressive guide-tree approach is its high tendency for error. The existing methods cause errors for two reasons. First, because the guide-tree merging order determined that other sequences were more distant, all alignment operations fail to consider any information outside of the sequence pair at large. For datasets with even a small degree of complexity, making the correct alignment requires exploiting relationships between all sequences. Second, using existing methods, an incorrectly placed activity cannot be corrected once it is aligned [4]. In this way, progressive guide-tree methods encourage alignment errors and do not allow correcting them.

Instead of using a progressive guide-tree, our algorithm uses an iterative approach that we have termed process-oriented iterative multiple alignment (PIMA). Iterative multiple sequence alignment is also widely used in bioinformatics [1][12], but has not been applied to process mining. By redesigning iterative alignment to better suit workflow data, we can achieve better scores than existing trace alignment algorithms while reducing the time complexity to $O(NL^2)$.

*C. Contribution*

- Process-oriented iterative multiple alignment (PIMA): an iterative multiple trace alignment framework specialized for workflow data. It outperforms existing progressive guide-tree methods, while significantly reducing time complexity by a factor of $O(N)$.
- Results from a medical case study using PIMA: we applied PIMA to workflow data obtained from the trauma resuscitation process. With this data, we demonstrate how our method can improve data representation and benefit knowledge extraction.

## II. METHODOLOGY

PIMA quickly creates an initial alignment, and then iteratively removes and realigns subsets of traces to the alignment matrix (Figure 1). We provide the pseudocode for PIMA (Alg.1); the steps are listed with corresponding section numbers and tight lower-bound time complexities.

| Algorithm 1. PIMA Framework | | |
|---|---|---|
| 0 | **T** ← given list of traces in log | |
| 1 | **G** ← build initialization guide tree    section II.B | $\Omega(N)$ |
| 2 | **M** ← **T** aligned by **G** | $\Omega(NL^2)$ |
| 3 | while **M** not converged:    section II.D | $\Omega(NL^2)$ |
| 4 |      **s** ← a subset of traces from **M**    section II.C | |
| 5 |      **s'** ← **M** – **s** | |
| 6 |      **M** ← alignment of **s** and **s'**    section II.A | $\Omega(L^2)$ |
| 7 | return **M** | |

As inferred from the time complexities of Alg.1, PIMA has a time complexity of $O(NL^2)$. This time is one magnitude $O(N)$ lower than the $O(N^2L^2)$ time required to calculate distance comparisons between each pair for edit-distance guide-trees. The time reduction is important for workflow analysis, as it is assumed that big-data workflow logs have a very large number of traces.

Without a metric-driven guide-tree to provide any direction for alignment merging order, PIMA incurs high penalties to sum-of-pairs score during the initialization (see initialization columns of Table I). By iteratively performing quick, small readjustments to the alignment matrix, it compensates for these errors and eventually outscores previous progressive guide-tree methods (see convergence columns of Table I).

PIMA is a flexible algorithm that is very open to variation. It is more of a general procedural framework that allows users to string specific pieces together for custom implementations. PIMA can be modified at almost every step of the process: initializing the alignment matrix, selecting subsets for realignment, and deciding on a convergence condition. Although many combinations of variations are possible, we will only focus on a few in this paper.

*A. Pair-Wise Alignment Operations*

All PIMA's alignments are performed using the Needleman-Wunsch dynamic programming algorithm [9], which takes $O(L^2)$ for a pair-wise alignment. To better suit the process mining application, we made several modifications. First, we assume that all activities have the same unit weight and do not allow activity substitution [4]. In other words, diagonal paths in the algorithm's matrix are permitted only if there is a match (last line of Eq. 3). All columns in the alignment matrix then consist of only one activity type. This restriction makes sense in the process mining context because activities are normally not quantifiably substitutable and because having multiple activities per column impedes interpretation of the final alignment matrix. If an application requires it, substitution can be implemented by preprocessing substitutable activities to have the same activity label.

Second, unlike previous alignment algorithms [4], we directly apply Needleman-Wunsch [9] to minimizing sum-of-pairs score (Eq. 1). This property may be achieved by taking advantage of the restriction we imposed on substitution and our unit activity weighing scheme. Because it is given that each column can only contain either gaps or activities of one type, each of the column's gaps incurs penalty from each of the column's activities and vice versa. The simplified sum-of-pairs score (Eq. 2) can be calculated in $O(NL)$ time.

$$SPS(M) = 2\sum_{k=1}^{L} f_k \times (N - f_k) \qquad (2)$$

Using this sum-of-pairs penalty scheme, we modify the Needleman-Wunsch objective function (Eq. 3), which is used in step 6 of the framework (Alg.1) (Figure 1).

$$Q_{i,j} = max \begin{cases} Q_{i,j-1} - 2g_j(N - g_j) \\ Q_{i-1,j} - 2f_i(N - f_i) \\ Q_{i-1,j-1} - 2(f_i + g_j)\big(N - (f_i + g_j)\big) \, if \, F_i = G_j \end{cases} \qquad (3)$$

**Q** is the matrix of values of the dynamic programming table [4]. The vectors **f** and **g** contain the number of activities in each of the columns of the two alignments being aligned, respectively. The vectors **F** and **G** hold the activity labels of each of the columns of the two alignments being aligned. **N** is the total number of traces.

The implication of using this objective function is that sum-of-pairs score will converge given enough iterations. This objective function ensures that sum-of-pairs score is monotonically non-increasing across iterations. Given an aligned log (step 2 in Figure 1), removing a subset of traces from the alignment matrix (step 4 and 5 in Figure 1) and realigning it to the rest of the log (step 6 in Figure 1) cannot increase sum-of-pairs score. In addition, the score is discrete and has a lower bound of zero (sum of Hamming distances cannot be negative). This is sufficient to prove that the sum-of-pairs score will converge given enough iterations. This fact is used in sections II.C and II.D.

There are some interesting notes about using our modified objective function (Eq. 3). Firstly, the value at the end of the alignment path in the dynamic programming table is exactly the negative sum-of-pairs score of the alignment. Additionally, observe that there is no positive reward given in any part of the algorithm; all values in the dynamic programming table are non-positive.

*B. Initialization*

As previously explained, PIMA's boost in computation time is due to its avoidance of computationally-expensive $O(N^2L^2)$ guide-trees. Instead, the initialization should aim to be quick, not necessarily accurate.

There are multiple initialization methods available. The simplest and least expensive is a random guide-tree, which takes $\Theta(N)$ to construct. One way to do this is to shuffle the traces and then align them one-by-one (i.e. align trace 1 to 2, then (1,2) to 3, then (1,2,3) to 4, etc.). We refer to this technique as "random sequential" merging (see Table I).

Performing such sequential merging considers the greatest number of traces per alignment operation on average. For example, aligning trace 1 to 2 considers two traces, (1,2) to 3 considers three, etc. It is common sense that more information given to the pair-wise alignment algorithm should improve the multiple alignment outcome. However, previous guide-tree methods usually do not merge in sequence as described, and so do not provide the optimal amount of information during alignment. Our experiments show that while the previous methods may outperform sequential merging initializations for small datasets, the previous method is outperformed by even random sequential initializations for larger logs (see initialization columns in Table I).

We also propose the second-cheapest tier of initialization guide-trees, which cost $O(NL+N \log N)$. These methods simply sort the traces by a simple metric and merge them one-by-one based on the metric's order. These are intended to provide a crude guide for building the tree, under the assumption that traces with similar metric values align better.

One eligible metric is the trace's number of activities (sorted length in Table I). Another somewhat more precise option is to arbitrarily assign each activity a unique positive number, and measure each trace by the sum of all its activity numbers (sorted activity sum in Table I). Sorting takes $O(N \log N)$ time and the metrics require $O(L)$ time per trace, so

TABLE I. PERFORMANCE COMPARISON OF PIMA RELATIVE TO EXISTING ALGORITHM.

|  | Intubation | | Trauma Resuscitation | | Primary Survey | | Artificial A | | Artificial B | |
|---|---|---|---|---|---|---|---|---|---|---|
|  | Initial. | Conv. | Initial. | Conv. | Initial. | Conv. | Initial. | Conv. | Initial. | Conv. |
| Previous (%) | 00.00 | -1.66 | 00.00 | -1.16 | 00.00 | -0.55 | 00.00 | -12.54 | 00.00 | **-16.50** |
| Sorted Act Sum (%) | 01.31 | -0.68 | 01.76 | -1.43 | 00.00 | -0.53 | -13.14 | **-13.76** | -12.98 | -14.70 |
| Sorted Length (%) | 02.30 | **-2.07** | 00.39 | **-1.71** | 00.00 | -0.53 | -11.15 | -12.25 | -15.37 | -15.88 |
| Rand. Sequential (%) | 00.22 ± 0.9 | -1.77 ± 0.5 | 01.01 ± 0.6 | -0.89 ± 0.3 | 00.28 ± 1.7 | **-0.64 ± 0.4** | -12.18 ± 0.9 | -12.39 ± 0.9 | -14.65 ± 1.0 | -15.01 ± 1.0 |

All values are percent change in SPS from respective scores using previous methods; negatives are improvements.

these O(NL+N log N) guide-trees still safely avoid the O($N^2L^2$) of existing algorithms. Our results demonstrate that sorted initialization may improve the converged alignment ("sorted" rows in Table I).

For small datasets, where N is negligible, the speed-up offered by PIMA becomes insignificant. In this case, it is more feasible to simply perform the initial alignment with a high-quality O($N^2L^2$) edit-distance guide-tree, and later use PIMA as a post-processor to correct alignment errors. Our results will demonstrate that PIMA indeed recovers errors made by the existing algorithm (see convergence of previous method in Table I). The ability to serve as a post-processor demonstrates PIMA's versatility and generalizability.

*C. Iteration*

The goal of iteration is to fix the errors incurred during initialization. PIMA achieves this by removing a subset of traces from, and then realigning it to, the alignment matrix (Alg.1) (step 4 and 6 in Figure 1). As explained in section II.A, all the alignment operations that occur during iteration are between a subset of the alignment and its complement, so all available information is exploited and sum-of-pairs score cannot increase.

Proper subset selection is crucial to PIMA's success. The easiest and most effective method is to choose a single trace: every iteration, each trace would be removed and realigned once. We refer to one cycle of this method as a single-trace realignment iteration (Alg.2). Our experiments show that the order of traces for realignment (step 1 of Alg.2) does not make a significant difference, so a random ordering may be used without loss of performance.

| Algorithm 2. Single-Trace Realignment (framework 4-6) | | |
|---|---|---|
| 0 | **M** ← given current alignment | |
| 1 | **R** ← given order of traces | Ω(N) |
| 2 | for **r** in **R**: | Ω($NL^2$) |
| 3 | **s'** ← **M**[**r**] | |
| 4 | **s** ← **M** − **M**[**r**] | |
| 5 | **M** ← alignment of **s** and **s'** | Ω($L^2$) |
| 6 | return **M** | |

As shown in section II.A, sum-of-pairs will converge to a minimum after sufficient iteration. As our results will show, it is likely that the resting point is a local minimum. We provide an example of such a local minimum (step 1 and 2 in Figure 1, nonoptimal alignment). Even though the best sum-of-pairs score can be achieved by aligning the light activities into a full column and splitting the darker ones (step 6 in Figure 1, optimal alignment), it is impossible to achieve the optimal alignment by iterating over only single traces. Whenever a removed trace is realigned, taking the correct step towards the right alignment would not be allowed by the "status quo" imposed by the rest of the alignment. To escape local minima, PIMA uses multi-trace realignment of subsets with size greater than one. In this paper, we introduce a simple multi-trace subset selection method (Alg.3).

The algorithm visualization is an example of a multi-trace realignment (step 4 and 6 in Figure 1). It shows one (s,s') pair in S being realigned (step 12 and 13 in Alg.3). We first filter out activity types that occupy only one column in the alignment matrix (step 2 in Alg.3). In this case, the grey activity's column is not considered (step 4 in Figure 1), because that grey column could not align to any other column. Among the remaining columns, we select ones with column frequencies between a certain range, typically around 10-90% or 20-80% full (step 1 and 7 in Alg.3). For example, a 20-80% threshold would remove both black columns from consideration (step 4 in Figure 1). These columns are either so full or so sparse that they likely cannot realign. Each of the final filtered columns then defines a candidate subset for multi-trace realignment; traces with an activity present in that column is in one subset, and those without it are in the other. In the presented example, the second column was the selected candidate for a subset split. The top subset (step 4 in Figure 1, upper) consists of traces containing activities in column two, while the bottom subset (step 4 in Figure 1, lower) has only gaps in column two. For every multi-trace realignment iteration, all candidate subset pairs are realigned in order of descending subset size (step 11 in Alg.3). In the example, only the realignment of column two was shown. For one multi-trace iteration, steps 4-6 would be repeated for columns 4, 7, 8, and 10 (step 4 and 6 in Figure 1). The optimal alignment with a lower sum-of-pairs score was achieved by removing and realigning this subset (step 6 in Figure 1, alignment path).

| Algorithm 3. Multi-Trace Realignment (framework 4-6) | | |
|---|---|---|
| 0 | **M** ← given current alignment | |
| 1 | **F** ← given range of column frequency required for candidacy | |
| 2 | **A** ← list of activities present in more than one column | Ω(L) |
| 3 | **S** ← empty list of subset pairs | |
| 4 | for column **m** in **M**: | Ω(L) |
| 5 | **a** ← activity label of **m** | |
| 6 | **f** ← activity frequency of **m** | |
| 7 | if **a** ϵ **A** & **f** ϵ **F**: | |
| 8 | **s** ← indices of traces with activities in **m** | |
| 9 | **s'** ← indices of traces without activities in **m** | |
| 10 | **S** ← **S** + (**s**,**s'**) | |
| 11 | **S** ← **S** sorted by **s** size | Ω(L log L) |
| 12 | for **s**,**s'** in **S**: | O($L^3$) |
| 13 | **M** ← alignment of **M**[**s**] and **M**[**s'**] | |
| 14 | return **M** | |

If multi-trace realignment results in a change in sum-of-pairs score, a lower minimum may become available. The new minimum can be found by repeatedly performing single-trace realignment until convergence occurs again. Our experiments show that performing multi-trace realignment after converging once with single-trace realignment puts the alignment on-track for a lower minimum at the second convergence (Table II) (see primary survey in Figure 2).

*D. Convergence*

The conditions for convergence can also be modified. As previously shown in section II.A, repeatedly performing single-trace realignment iterations will eventually make sum-of-pairs score converge. Sum-of-pairs can be calculated in O(NL) time (Eq. 2), so using it as a convergence indicator every iteration should not have a noticeable effect on run time. For relatively small datasets of simple processes with few activities, sum-of-pairs score rapidly converges in two or three iterations after initialization (Figure 2).

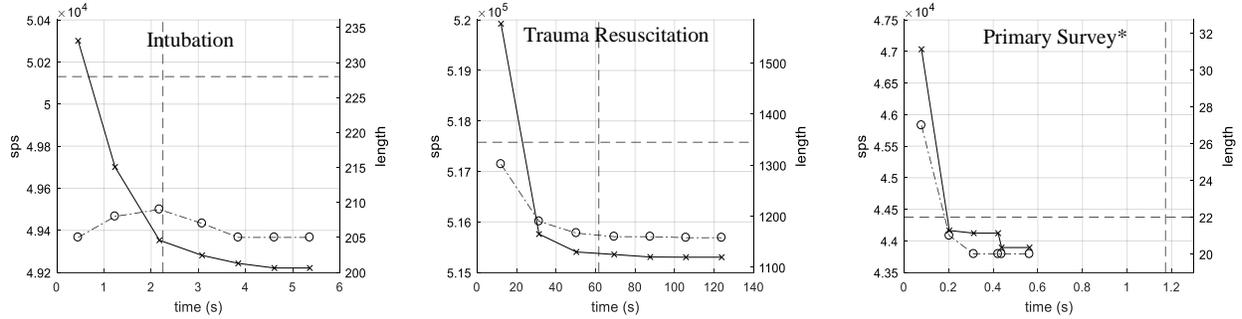

Figure 2. SPS and alignment length with respect to time for 3 real datasets. Each point is an iteration; the first point is initialization (random sequential method). Solid x-marked line is SPS; dashed o-marked line is length. The intersection of the two dashed lines represents the previous method results: the vertical is its time; the horizontal is scaled to both its SPS and length. *The Primary Survey converged at iteration 3; the sharp drop is a multi-trace realignment iteration.

The opposite also applies to complex datasets: it may take more than a dozen iterations to consecutively achieve the same sum-of-pairs score. Our experiments have shown that most changes in the alignment occur between initiation and the first iteration (Table III). Even for complex datasets, the sum-of-pairs score improvement per iteration usually does not exceed 1%, indicating that the matrix has already approached its local minimum. If the user only needs a rapid data visualization and does not require the absolute best accuracy, it may be more economical to terminate the iteration when the improvements begin to diminish (e.g., sum-of-pairs score reduction becomes under 1%).

## III. EXPERIMENTAL RESULTS

This section quantitatively assesses PIMA relative to the existing edit-distance guide-tree methods. We first describe our datasets, and then present our experimental findings. All experiments were performed on an Intel i5-4590 3.30GHz CPU with 16GB RAM, and all algorithms were implemented using the Anaconda distribution of Python 3.6.

### A. Description of Data

For our evaluation of PIMA, we used both real-life and artificial logs (Table IV). The first three datasets (endotracheal intubation, primary survey, and trauma resuscitation) were obtained by manually coding activity timestamps from videos taken at CNMC, a Level 1 trauma center in Washington, D.C. Use of these data was approved by the hospital's Institutional Review Board. The fifth dataset was collected at a Dutch hospital and is publicly available [8]. Because of the enormous size and complexity of the dataset (Table IV), we used it to demonstrate PIMA's computational capabilities on a large dataset. As all this data was collected by observing real processes, the relationships between activities in the traces are very complex, and the logs are very noisy.

The artificial logs (Table IV) were generated by the Process Log Generator v2 [7]. In this paper, we use the artificial log data only to demonstrate PIMA's effectiveness on large datasets. We adjusted the settings to produce minimal noise in the artificial logs.

### B. Time Complexity

We first examine the time advantage PIMA has over the progressive edit-distance guide-tree method (Table III). For each dataset, we ran and timed 30 random PIMA initializations (using the sequential merging method) and performed single-trace iterations until they converged. We then performed the existing alignment algorithm and recorded its run time. We show the time of the previous method, followed by the average time it took for PIMA to surpass its sum-of-pairs score, as well as the average number of iterations

TABLE II. EFFECTS OF MULTI-TRACE REALIGNMENT AFTER CONVERGING ONCE.

|  | Intubation | Trauma Resuscitation | Primary Survey | Artificial A | Artificial B |
|---|---|---|---|---|---|
| Immediate | -0.13 ± 0.55% | -0.23 ± 0.30% | -0.40 ± 0.10% | -0.74 ± 0.70% | -1.96 ± 0.73% |
| After Converging | -0.16 ± 0.55% | -0.52 ± 0.30% | -0.44 ± 0.00% | -0.96 ± 0.56% | -2.12 ± 0.72% |

Values shown are SPS percent change relative to before multi-trace realignment (with 10-90% thresholds); negatives are improvements. The immediate effect is shown, followed by the new converged minimum.

TABLE III. PIMA TIME COMPLEXITY COMPARISON WITH PREVIIOUS METHOD.

|  | Intubation | Trauma Resuscitation | Primary Survey | Artificial A | Artificial B |
|---|---|---|---|---|---|
| Previous (s) | 1.81 | 48.62 | 1.16 | 1094.26 | 3341.99 |
| Time to Beat Existing (s) | 0.74 ± 0.43 | 24.07 ± 6.52 | 0.11 ± 0.06 | 11.08 ± 0.44 | 18.43 ± 9.42 |
| Speedup | **x2.46** | **x2.02** | **x10.52** | **x98.75** | **x181.35** |
| Time/Iteration (s) | 0.55 ± 0.08 | 12.83 ± 2.59 | 0.10 ± 0.02 | 13.23 ± 2.22 | 21.39 ± 9.28 |
| Iterations to Beat Previous | 0.62 ± 0.72 | 1.00 ± 0.45 | 0.32 ± 0.47 | 0.00 ± 0.00 | 0.00 ± 0.00 |
| Failed to Beat Previous* | 3.33% | 0.00% | 6.67% | 0.00% | 0.00% |
| Failed to Beat Previous by | 0.10 ± 0.00% | - | 0.41 ± 0.05% | - | - |

*For all instances where PIMA failed to beat the previous method, performing one multi-trace realignment iteration allowed PIMA to surpass the existing algorithm.

it took (initialization is iteration zero). We include the percentage of instances where PIMA failed to outscore the previous method, as well as the average relative score difference across each case. In addition, we also calculated the average iteration time for each dataset.

PIMA outscores the previous method in less than half the time (Table III). Especially for the larger datasets, PIMA results in over 100-fold faster performance. For initializations that lead to better minima, PIMA usually does not take more than one single-trace iteration to outperform the existing algorithm. The average number of iterations to outscore the edit-distance guide-tree method is usually under one, indicating that many PIMA initializations already perform better than the previous methods. For the few cases where PIMA does not outperform the existing algorithm within one convergence, results using PIMA are still competitive (within half a percent), and PIMA was always able to surpass the previous method by performing one multi-trace iteration.

For the three clinical datasets, we graphically represent the performance and run time of PIMA with respect to the existing algorithms (Figure 2). For average-size real-world datasets, these are typical examples of what one should expect from PIMA. These charts confirm the intuitions underlying PIMA: it first rapidly builds an initialization alignment, and then iteratively corrects errors until convergence, eventually surpassing the score of previous methods while saving time (Figure 1). While PIMA may take longer to converge to the absolute minimum of a given initialization, these results demonstrate that PIMA has likely already outperformed the previous method within one iteration (Table III) (Figure 2). The rest of the time spent on convergence should be regarded as fine-tuning the alignment and can be stopped at any point as the user sees fit.

The Dutch hospital dataset is a special case. Because of its enormous size (Table IV), we display the results separately (Table V). The most important result of the Dutch hospital experiments was that the existing algorithm could not deliver an alignment within a reasonable time. We halted the program after 24 hours, and then approximated the total time to take more than 14 days. In contrast, PIMA was able to provide an initial alignment (sorted length method) and perform several iterations in just over 5 hours. Because the previous method failed to feasibly return any result at all illustrates the limitation of having an $O(N^2)$ term in the time complexity. From a computational standpoint, PIMA is better suited to handle large, complex datasets.

*C. Sum-of-pairs Score Performance*

We compared the different initializations of PIMA with respect to the previous progressive edit-distance guide-tree

TABLE V. DUTCH HOSPITAL DATASET EXPERIMENTAL RESULTS.

| It. | It. Type | SPS | Length | Time (s) | Align. Ops | Time/Op |
|---|---|---|---|---|---|---|
| 0 | rand. seq. | 96153730 | 21794 | 2947 | 833 | 3.54 |
| 1 | single | 94678604 | 18433 | 6813 | 833 | 8.18 |
| 2 | single | 94586140 | 18002 | 6142 | 833 | 7.37 |
| 3 | multi | 94580316 | 17987 | 2268 | 9 | 252.00 |

The existing algorithm failed to return an alignment within 24 hours; it would take approximately 14 days to complete the computations.

method (Table I). All values are the percent differences in sum-of-pairs score relative to the results from existing methods, with negative percentages indicating improvement. For each dataset, we list the relative sum-of-pairs score for each initialization method at initialization, followed by the relative score achieved after converging once from single-trace iteration. The values shown for the random initializations are averaged across thirty different seeds, and are shown with their standard deviations.

These results reveal several insights. First, PIMA outperforms the previous guide-tree methods, as indicated by all non-negative values at the first convergence (Table I). Because of the larger sizes of the artificial datasets, the percentage improvement from the previous method is more evident. PIMA was always able to improve from the initial alignment, even if the alignment was initialized using the existing algorithm.

Another observation is that the results from the existing algorithm are very poor. For the real datasets, using an edit-distance guide-tree does not offer a significant score increase compared to much more rapid guide-trees, including those that are randomly constructed. For the larger datasets, using an edit-distance guide-tree even negatively impacts the alignment quality, demonstrating that the edit distance metric is unable to properly guide the alignment of more than one hundred traces at a time, posing an issue for big data analysis. The poor performance of the previous methods also shows the disadvantage of performing alignments without factoring in the maximum amount of log information.

We also show the effects of multi-trace realignment (Table II). For each dataset, thirty random initializations were performed (random sequential method) and led to convergence; the values shown are averages with standard deviations across those thirty initializations. The thresholds we used to filter out candidate columns was 10-90%. The table shows the sum-of-pairs score of alignments immediately after the operation and at the resulting new converged minimum, all relative to the score of the minimum prior to the operation. Again, negative percentages are improvements. In addition, for the primary survey dataset, we graphically show the

TABLE IV. DATASET CHARACTERISTICS.

| | Intubation | Trauma Resuscitation | Primary Survey | Secondary Survey* | Dutch Hospital | Artificial A | Artificial B |
|---|---|---|---|---|---|---|---|
| Num. of Traces | 74 | 122 | 171 | 122 | 833 | 1000 | 1000 |
| Total Activities | 1293 | 7153 | 831 | 3057 | 147830 | 41548 | 73429 |
| Activity Types | 19 | 44 | 5 | 17 | 249 | 57 | 88 |
| Avg. Trace Length | 17.47 | 58.63 | 4.86 | 25.06 | 177.47 | 41.47 | 73.47 |
| Trace Length STD | 7.00 | 15.55 | 0.38 | 8.01 | 217.81 | 2.59 | 2.31 |

*The secondary survey is a shortened version of the trauma resuscitation dataset; it is only used for the medical case study (section IV).

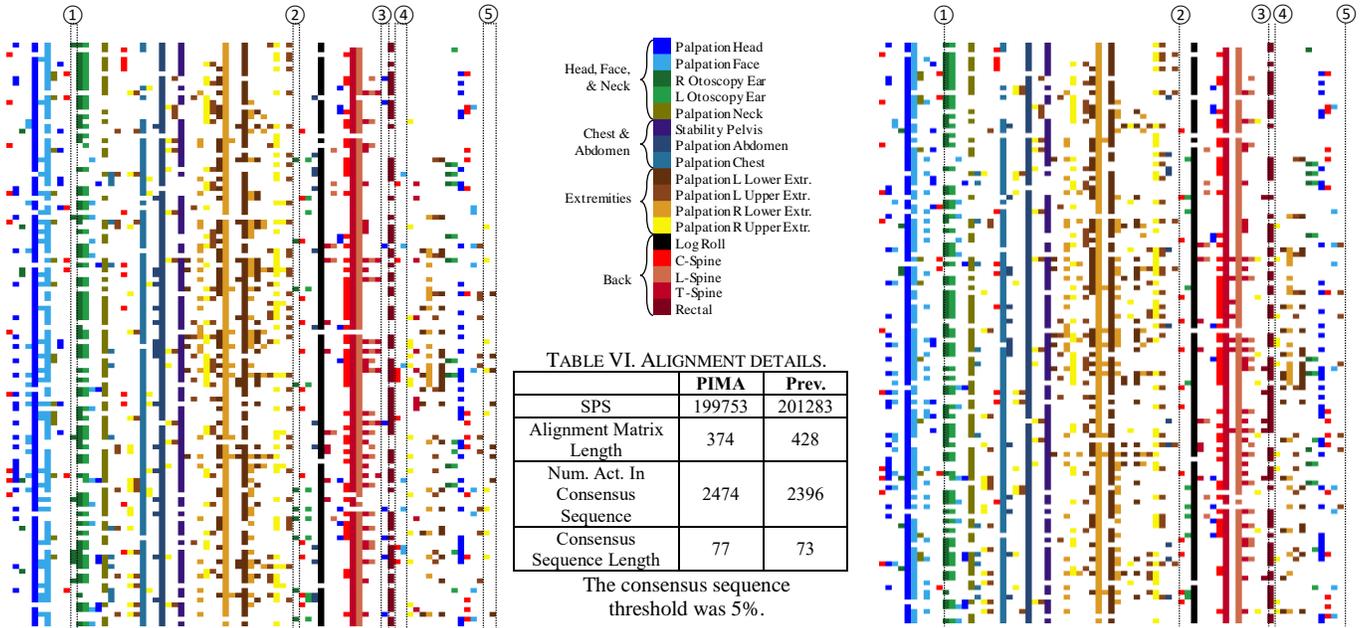
Figure 3. Comparison between filtered consensus sequence alignment results from PIMA (left) vs. the existing algorithm (right). (Best viewed in color.)

potential of multi-trace realignment (primary survey iteration 4 in Figure 2).

Performing multi-trace realignment iterations does move the alignment closer to the global minimum. In addition, the experiments confirm that the modified objective function (Eq. 3) makes sum-of-pairs score monotonically non-increasing, even for the alignment of more than one trace at a time. This direct minimization of sum-of-pairs score applies to datasets of any size, as shown by the decreasing sum-of-pairs score in the Dutch hospital experiments (Table V).

## IV. MEDICAL CASE STUDY

We performed a case study to more qualitatively compare PIMA with the existing methods and gain insights from our medical data. We evaluated how well the new alignment represents the data, and how it positively affects the extraction of knowledge. For the purposes of this paper, we only perform analysis on the secondary survey of the trauma resuscitation dataset, which contains 122 traces with 3057 activities of 17 types (Table IV). The secondary survey occurs towards the end of the resuscitation, and consists of examination activities categorized by location on the body (Figure 3 legend).

We initialized PIMA randomly, and then performed single-trace and multi-trace realignment iterations until sum-of-pairs score converged. The final sum-of-pairs score was achieved after 14 iterations, and is lower compared to that of the previous algorithm (Table VI). This result complements the shorter length of the PIMA alignment (Table VI). These observations quantitatively indicate that PIMA has found better ways of condensing common activities from different traces into more succinct columns with higher frequencies.

Before visual analysis of the alignments, we first perform the common practice of taking the consensus sequence [13]. The consensus sequence is a filtered version of an alignment, with columns of frequency less than a certain threshold removed. Visualizing the consensus sequence reduces the clutter caused by infrequent or insignificant activity columns. The frequency threshold is different for each dataset. For our data, a 5% minimum frequency was used to obtain a reasonable visualization (Figure 3). The properties of the consensus sequence are also indicative of the alignment's quality. Longer consensus sequences with more activities are preferable, indicating that the alignment algorithm better consolidated common activities into columns, resulting in more columns with frequencies high enough to surpass the consensus sequence thresholding. Compared to the consensus sequence from the previous method, PIMA's consensus sequence had both more activities and longer length (Table VI). Due to its better performance, the PIMA consensus sequence contained 78 activities and 4 columns that would have otherwise been omitted from visualization (Table VI).

The additional columns in the PIMA consensus sequence also allow for higher-quality knowledge extraction. In the visualization, we label a few medically significant columns in PIMA's consensus sequence (Figure 3 dashed boxes) that are not present in the visualization of the previous method (Figure 3 dashed lines). The first and fifth boxes are right ear otoscopies and extremity palpations, respectively (①⑤ in Figure 3). The columns may not be very significant from a medical standpoint, but they give a better sense of the trauma resuscitation's nature. At a high level, these columns help convey that the process is complex and flexible: some activities may or may not be repeated at different times in the workflow, and the performers might not strictly adhere to having a specified number of activity occurrences.

The second box also contains right ear otoscopies (② in Figure 3). Because this column is missing from the consensus sequence of the existing method, the data visualization fails to convey that delayed left otoscopies (Figure 3, two columns after ②) are still paired with a right otoscopy. Because PIMA properly consolidates activities, the right-left ear pairing is present in the PIMA consensus sequence. The presence of the pairing highlights PIMA's ability to better represent the data through alignment. Otoscopy (Figure 3, green) is expected to occur during the head exam early on in the secondary survey. The alignment, however, visually conveys that otoscopy can occur after the extremity exams or even as late as during the back exam. The left and right otoscopies still usually occur together, regardless of their location in the overall process.

The third box and second column of the fourth box (③④ in Figure 3) are head and face palpations. Based on our medical domain knowledge, we did not anticipate the presence of this many head and face palpations (Figure 3, light blue) during the back exam (Figure 3, reds). Further investigation showed that the team often manipulated the patient's head when rolling the patient (Figure 3, black) to avoid excessive movement of the neck. We thus adjusted our expectations to allow late head and face palpations.

Lastly, we examine the first column of the fourth box (④ in Figure 3), which contains cervical-spine exams. In the trace log, the cervical spine exam occurs at least once in 87.7% of cases, making it an important activity in the secondary survey. On average, it only occurred just over one time per case, meaning it is not often repeated. A trace visualized without the cervical spine activity would then appear deviant. By not having the indicated column, the consensus sequence of the previous method leaves out 5.0% of the cervical-spine activities in the log, reducing the number of traces with at least one spine exam by 4.1%. The failure of the previous method to align sufficient cervical spine activities increases the number of wrongly-perceived deviations in the visualization. PIMA, on the other hand, correctly aligned these activities and presented them in its consensus sequence.

From the medical perspective, cervical-spine exams should occur with the rest of the back exam. In the data visualization (Figure 3, bright red), however, cervical-spine exams often occur during the head exams. Further investigation revealed that cervical-spine exams were performed early alongside neck-related activities so that the medical team could determine whether the patient required cervical-spine stabilization. It is clinically acceptable to perform the cervical-spine exam during the head and neck exams instead of during the back exams.

This case study contains just a few examples of how alignment can aid workflow analysis, and demonstrates how PIMA's better quantitative score translates to improved qualitative analysis.

## V. CONCLUSION

While PIMA has many advantages over the previous progressive guide-tree algorithms, it still has drawbacks. PIMA still might not find the global minimum. Multi-trace realignment also incurs heavy computational penalties when handling extremely sparse alignments. Better column candidacy criteria and more subset selection methods are needed. Also, heuristic data-driven rules should be developed to determine the method most likely to be successful.

We proposed the novel process-oriented iterative multiple alignment (PIMA) framework, which is the first iterative trace alignment algorithm used in the process mining context. Using optimizations for large workflow datasets, PIMA is able to outscore previous progressive guide-tree alignment methods on actual and artificial data while significantly reducing the time complexity by a factor of O(N). We applied alignment to analyzing a medical process, showing how PIMA can better represent workflow data and facilitate the extraction of insights. We hope to have shown the process mining community the benefits of iterative alignment.